\documentclass[twocolumn,english]{IEEEtran}
\usepackage[T1]{fontenc}
\usepackage{color}
\usepackage{babel}
\usepackage{array}
\usepackage{amsmath}
\usepackage{amssymb}
\usepackage{graphicx}
\usepackage[unicode=true,
 bookmarks=true,bookmarksnumbered=true,bookmarksopen=true,bookmarksopenlevel=1,
 breaklinks=false,pdfborder={0 0 0},pdfborderstyle={},backref=false,colorlinks=false]
 {hyperref}
\hypersetup{pdftitle={Your Title},
 pdfauthor={Your Name},
 pdfpagelayout=OneColumn, pdfnewwindow=true, pdfstartview=XYZ, plainpages=false}
\usepackage{breakurl}

\makeatletter

\providecommand{\tabularnewline}{\\}

\ifCLASSOPTIONcompsoc
\usepackage[caption=false,font=normalsize,labelfont=sf,textfont=sf]{subfig}
\else
\usepackage[caption=false,font=footnotesize]{subfig}
\fi
\usepackage{cite}
\usepackage{amsthm}

\usepackage{algorithm}
\usepackage{amsmath}

\@ifundefined{showcaptionsetup}{}{%
 \PassOptionsToPackage{caption=false}{subfig}}
\usepackage{subfig}
\makeatother

\begin{document}

\title{Network Densification in 5G: From the Short-Range Communications
Perspective}

\author{\IEEEauthorblockN{Junyu Liu, Min Sheng, Lei Liu, Jiandong Li}\\
\IEEEauthorblockA{State Key Laboratory of Integrated Service Networks,
Xidian University, Xi'an, Shaanxi, 710071, China\\
Email: junyuliu@xidian.edu.cn, \{msheng, jdli\}@mail.xidian.edu.cn,
leiliu@stu.xidian.edu.cn}}
\maketitle
\begin{abstract}
Besides advanced telecommunications techniques, the most prominent
evolution of wireless networks is the densification of network deployment.
In particular, the increasing access points/users density and reduced
cell size significantly enhance spatial reuse, thereby improving network
capacity. Nevertheless, does network ultra-densification and over-deployment
always boost the performance of wireless networks? Since the distance
from transmitters to receivers is greatly reduced in dense networks,
signal is more likely to be propagated from long- to short-range region.
Without considering short-range propagation features, conventional
understanding of the impact of network densification becomes doubtful.
With this regard, it is imperative to reconsider the pros and cons
brought by network densification. In this article, we first discuss
the short-range propagation features in densely deployed network and
verify through experimental results the validity of the proposed short-range
propagation model. Considering short-range propagation, we further
explore the fundamental impact of network densification on network
capacity, aided by which a concrete interpretation of ultra-densification
is presented from the network capacity perspective. Meanwhile, as
short-range propagation makes interference more complicated and difficult
to handle, we discuss possible approaches to further enhance network
capacity in ultra-dense wireless networks. Moreover, key challenges
are presented to suggest future directions.
\end{abstract}

\section{Introduction\label{sec:Introduction}}

\IEEEPARstart{I}{nstead} of the enhancement of radio access networks
(RANs), the future wireless networks are more likely to act as a mixture
of various types of RANs, e.g., macro-cell base stations (BSs), femto-cell
BSs, pico-cell BSs and WiFi access points (APs), etc. By 2030, the
targets are to \textcolor{black}{support ubiquitous device connectivity
and expand network capacity}, including 100 billion device connections,
20000$\times$ mobile data traffic and 1000$\times$ user experienced
data rate, etc., compared to 2010 \cite{5G_Requirement_Ref}. The
requirements are even more critical for popular scenarios. For instance,
tens of Tbps/$\mathrm{km}^{2}$ is required in the office, 1 million
connections/$\mathrm{km}^{2}$ is required at densely populated areas
such as stadium and open gathering, and super high density of over
6 persons/$\mathrm{m}^{2}$ is to be supported in subways. Among the
appealing approaches to realize the ambitious goals, network densification
is shown to be the most promising one, which has improved network
capacity by 2700 folds from 1950 to 2000 \cite{UDN_benefit_ref}.
The principle of network densification is to deploy BSs/APs with smaller
coverage and enables local spectrum reuse. In consequence, users can
be served with shorter transmission links, thereby fully exploiting
spatial and spectral resources.

However, does aggressively deploying more BSs and devices always improve
the system performance? As network densification significantly reduces
transmission distance and enables proximity communication, the signal
propagation may transit from long- to short-range propagation. For
instance, as shown in Table \ref{Table 1}, if the BS density is increased
from 1 $\mathrm{BS}/\mathrm{km}^{2}$ to 100 $\mathrm{BS}/\mathrm{km}^{2}$,
the average transmission link length is reduced 10 folds from 500
m to 50 m, which makes a larger proportion of downlink users located
within the short-range propagation distance. Meanwhile, the application
of device-to-device (D2D) communications allows direct local transmission
of nearby users to bypass centralized BSs, which greatly shortens
the distance between transmitters (Tx's) and the potential receivers
(Rx's) as well. As will be discussed later, signal propagation features
are different within long- and short-range regions, e.g., signal strength
decays moderately with transmission distance in short-range regions,
while decays more quickly in long-range regions. Accordingly, signal
attenuation features have to be modeled differently. However, in traditional
wireless communications system, the performance evaluation and protocol
design are basically based on the assumption of long-range transmission.
Therefore, the impact of network densification and short-range transmission
on network performance remains to be explored.

Recently, the research of densely deployed wireless networks has gained
great attention from both academia and industries, including architecture
development \cite{Architecture_Ref,User_Centric_UDN_Ref}, analytical
studies \cite{Ref_multi_slope_1,Ref_multi_slope_2} and protocol design
\cite{Resource_allocation_Ref}. Remarkably, a user-centric architecture
has been presented for ultra-dense networks (UDN) \cite{User_Centric_UDN_Ref},
under which functions such as mobility management, resource management
and interference management, can be co-designed and jointly optimized.
Meanwhile, to comprehensively understand the merits and limits brought
by network densification, authors study the network capacity scaling
law in downlink cellular network\cite{Ref_multi_slope_1,Ref_multi_slope_2}.
Wherein, the impact of both line-of-sight (LoS) and non-LoS (NLoS)
transmissions has been fully explored. Depending on parameters, it
is shown that the spatial throughput, which is an indicator of network
capacity, scales slowly and even diminishes to be zero when BS density
is sufficiently large \cite{Ref_multi_slope_1}. The above results
indicate that network densification may be beneficial, while network
ultra-densification would lose the merits of network capacity enhancement
when spatial resources are fully exhausted. Nevertheless, despite
the progress achieved by recent research, there is still no consensus
on how dense is ultra-dense in wireless networks. More importantly,
the available study fails to capture the influence of short-range
transmissions in UDN, which makes the existing results dubious and
doubtful.

With this regard, we intend to characterize ultra-densification for
wireless networks by fully exploring short-range propagation features.
To this end, we first provide answers to the following two questions:
1) how to characterize short-range propagation features and 2) how
short-range propagation influences the performance of wireless networks
in terms of spatial throughput. Aided by the spatial throughput scaling
law, we then concretely reveal how dense is ultra-dense in wireless
networks. To combat interference and further enhance the system performance
in UDN, we overview the state-of-the-art technologies like interference
management, non-orthogonal multiple access (NOMA), and millimeter-wave
(mm-Wave) communications, and key challenges to facilitate them in
UDN are highlighted as well.

The rest of this article is organized as follows. We first discuss
signal propagation features within short distance via experimental
results. Then, the interpretation of ultra-densification is presented
from the spatial throughput scaling perspective. Following that, the
detail of the possible approaches to further enhancing the capacity
in UDN is discussed, and open challenges brought by ultra-densification
are provided as well. Finally, conclusion remarks are drawn.

\section{Short-Range Propagation in UDN}

\begin{table*}[t]
\begin{centering}
\begin{tabular}{|>{\centering}p{2cm}|>{\centering}p{2cm}|c|>{\centering}p{3cm}|>{\centering}p{3cm}|>{\centering}p{3cm}|}
\hline
{\footnotesize{}Avg. of link length $d$ (m)} & {\footnotesize{}BS density $\left(\mathrm{BS}/\mathrm{km}^{2}\right)$} & {\footnotesize{}$\mathtt{P}\left(d<50\:\mathrm{m}\right)$} & {\footnotesize{}$\mathtt{P}\left(d<20\:\mathrm{m}\right)$} & {\footnotesize{}$\mathtt{P}\left(d<10\:\mathrm{m}\right)$} & {\footnotesize{}$\mathtt{P}\left(d<1\:\mathrm{m}\right)$}\tabularnewline
\hline
\hline
{\footnotesize{}500} & {\footnotesize{}1} & {\footnotesize{}0.78\%} & {\footnotesize{}0.13\%} & {\footnotesize{}0.03\%} & {\footnotesize{}$3.1\times10^{-6}$}\tabularnewline
\hline
{\footnotesize{}100} & {\footnotesize{}25} & {\footnotesize{}17.8\%} & {\footnotesize{}3.1\%} & {\footnotesize{}0.78\%} & {\footnotesize{}0.01\%}\tabularnewline
\hline
{\footnotesize{}50} & {\footnotesize{}100} & {\footnotesize{}54.4\%} & {\footnotesize{}11.8\%} & {\footnotesize{}3.1\%} & {\footnotesize{}0.03\%}\tabularnewline
\hline
{\footnotesize{}10} & {\footnotesize{}2500} & {\footnotesize{}$\approx$100\%} & {\footnotesize{}95.7\%} & {\footnotesize{}54.4\%} & {\footnotesize{}0.8\%}\tabularnewline
\hline
{\footnotesize{}5} & {\footnotesize{}10000} & {\footnotesize{}$\approx$100\%} & {\footnotesize{}$\approx$100\%} & {\footnotesize{}95.7\%} & {\footnotesize{}3.1\%}\tabularnewline
\hline
\end{tabular}
\par\end{centering}
\vspace{0.2cm}

Table I. \label{Table 1}The probability that cellular transmission
occur within different distances. The above statisitics are obtained
via simulations provided that users are connected to the geometrically
nearest BSs.
\end{table*}

It is evident that network densification would push Tx's and Rx's
closer to each other. Taking cellular networks as an example, when
small cells are deployed for ubiquitous and seamless coverage, the
transmission distance from users to small cell BSs is greatly reduced
from hundreds of meters (macro cell case) to tens of meters. In the
upcoming fifth generation (5G) wireless networks, small cell BSs of
plug-and-play features would be deployed anywhere (on the tables or
the roof of a room) and accordingly the transmissions would occur
within much shorter distance (e.g., up to 10m). For instance, when
the \textit{nearest neighbor connectivity rule} is adopted, i.e.,
users are supposed to connect to the geometrically closest BSs, the
probabilities that transmission occurs within different regions are
shown in Table I. It is observed that the probability that transmission
occurs within 20m raises 91 folds from 0.13\% to 11.8\% when BS density
increases from 1 $\mathrm{BS}/\mathrm{km}^{2}$ to 100 $\mathrm{BS}/\mathrm{km}^{2}$.
If BS density further increases from 1 $\mathrm{BS}/\mathrm{km}^{2}$
to 2500 $\mathrm{BS}/\mathrm{km}^{2}$, this probability astonishingly
raises 750 folds from 0.13\% to 95.7\%. This is especially true for
scenarios like offices and dining halls, etc. For more crowded places,
such as stadiums, more dense BS deployment is required such that transmission
distance is further reduced. In addition, we see from Table I that
almost all the transmissions occur within 10 m when the BS density
reaches 10000 $\mathrm{BS}/\mathrm{km}^{2}$.

It is worth noting that signal propagation features within short distance,
e.g., up to 10m, significantly differ from those in the long distance,
e.g., larger than 50m. Even when they are located in proximity and
LOS path exists between them, signal propagations may be still different.
For instance, when Tx's are dozens of meters apart from Rx's, the
LOS component of signals dominates the NLOS component and accordingly,
the wireless channels between them are likely to experience Rician
fading. In contrast, when Tx's and Rx's are in close proximity, the
LOS and NLOS components become less distinguishable. As will be shown
later, it is more likely that the channel between Tx's and Rx's is
a Rayleigh fading channel.

According to the above discussion, to better understand the impact
of network densification, it is essential to figure out signal propagation
features over different propagation distance. To shed light on this,
we discuss short-range propagation features via experimental results
in the following.

\begin{figure*}[t]
\begin{centering}
\subfloat[\label{fig:Experimental scenarios}Experimental scenarios.]{\begin{centering}
\includegraphics[width=2in]{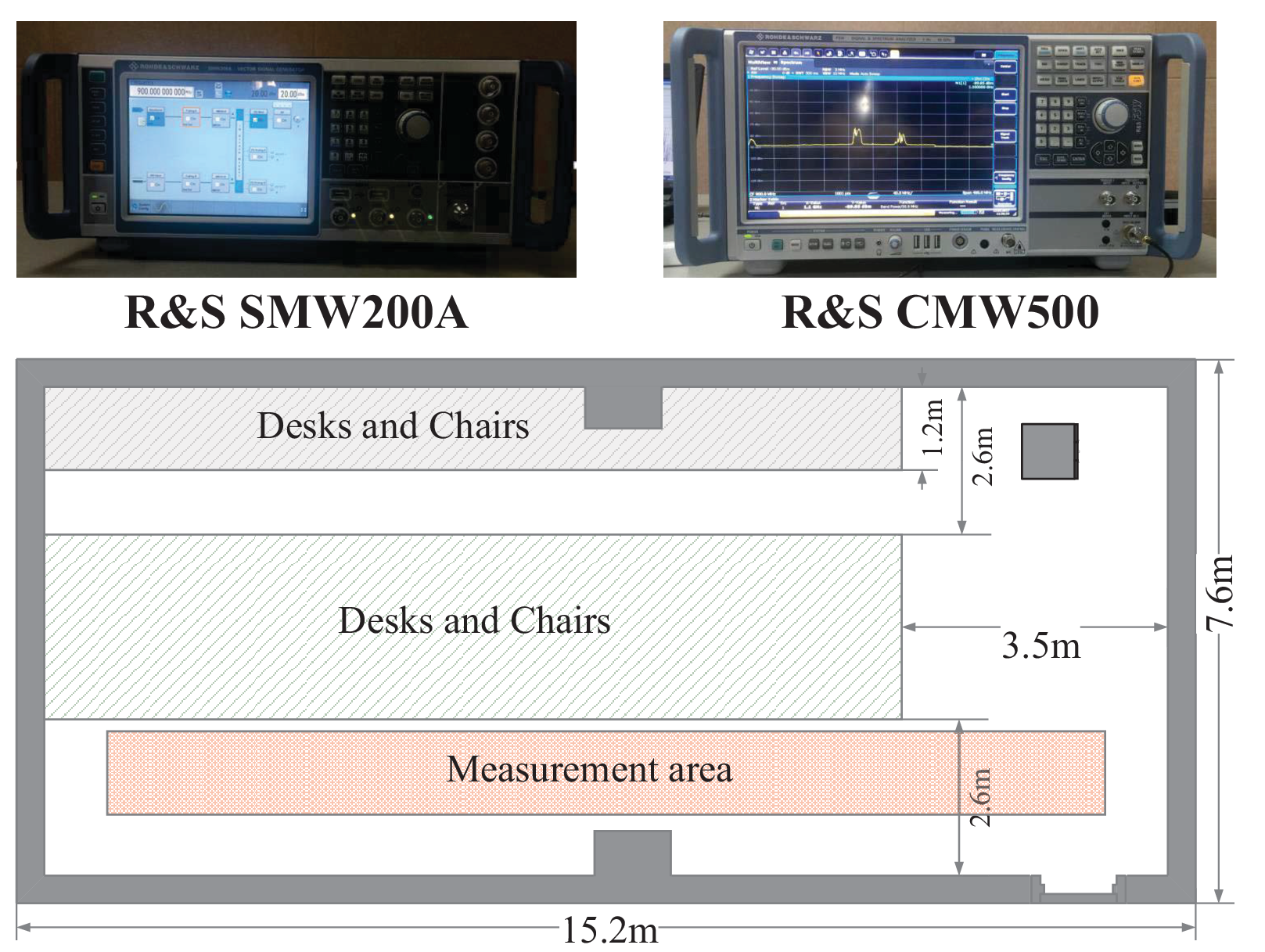}
\par\end{centering}
}\subfloat[\label{fig:Real experimental scenarios}Experimental setups.]{\begin{centering}
\includegraphics[width=2in]{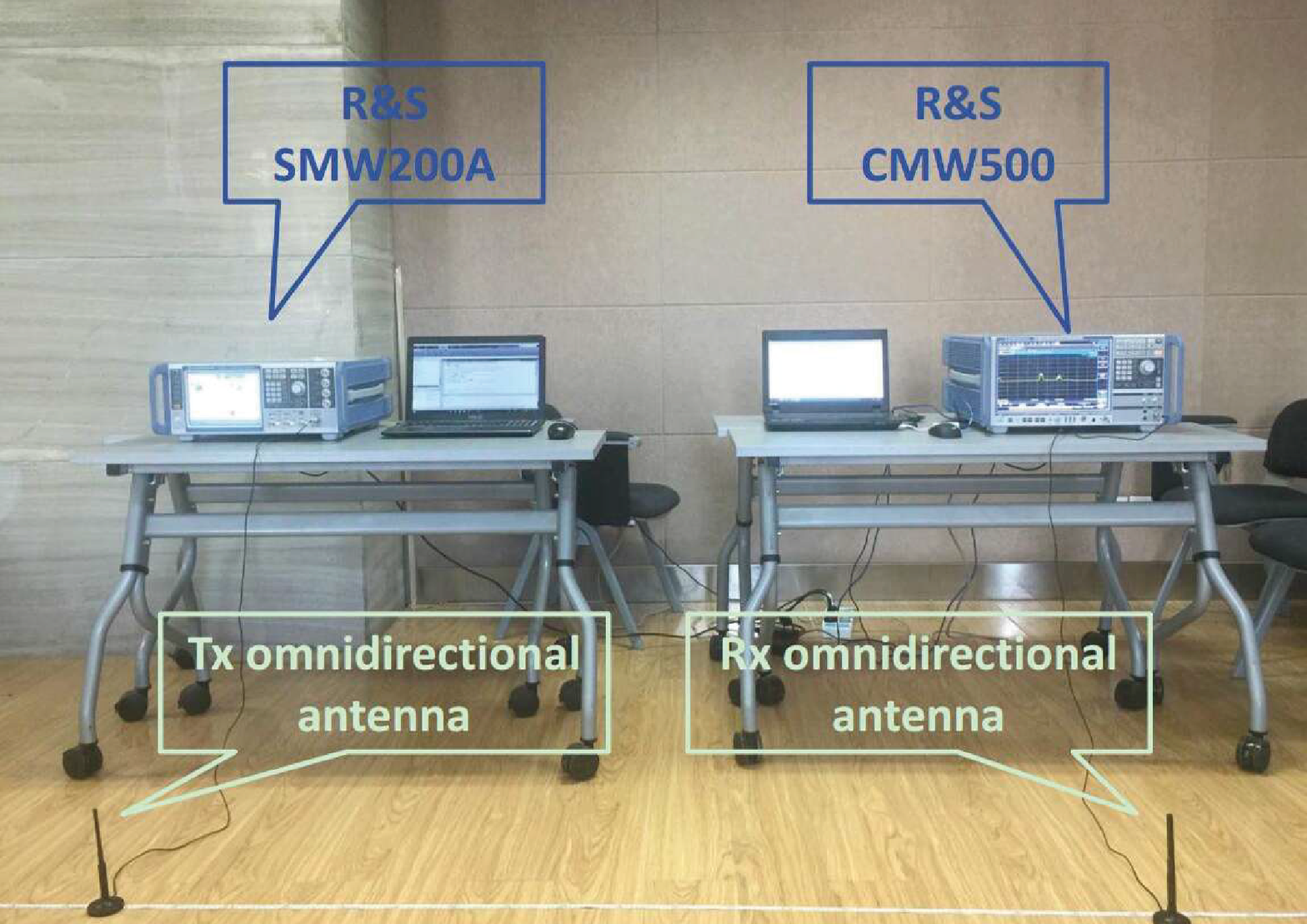}
\par\end{centering}
}\subfloat[\label{fig:Experimental and fitting results}Experimental and fitting
results.]{\begin{centering}
\includegraphics[width=2.3in]{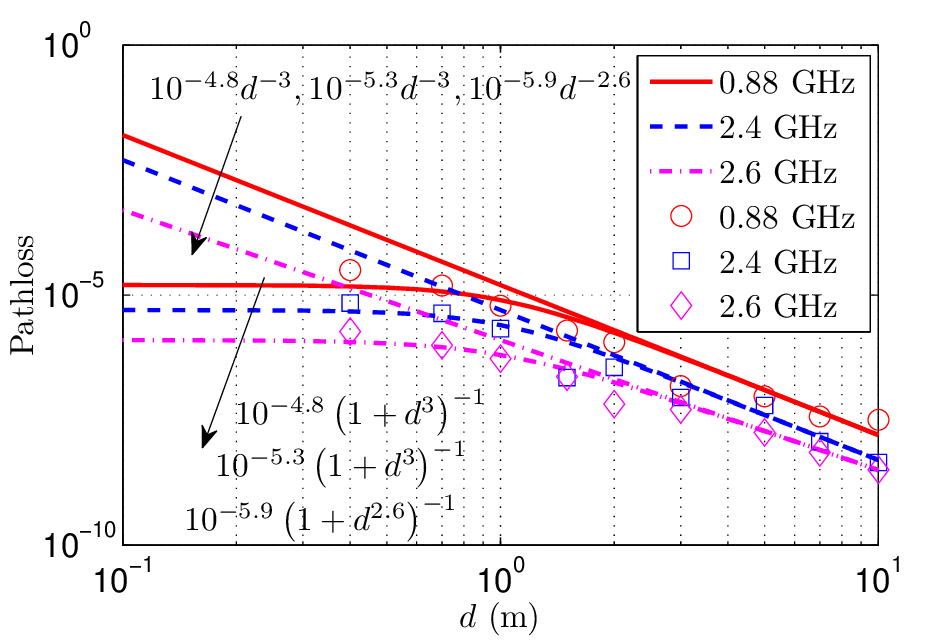}
\par\end{centering}
}
\par\end{centering}
\caption{\label{fig:Experimental results}Experimental results of received
signal power for short-range transmission. In (b), markers denote
the experimental results and lines denote the fitting results. Transmit
power of the signal generator is set to be 0 dBmW.}
\end{figure*}

\subsection{Experimental Results of Short-Range Signal Propagation}

In this part, we use experimental results to illustrate how pathloss
varies with transmission distance $d$ especially when $d$ is small.
The measurement is conducted in the meeting room of size 15.2m$\times$7.6m,
as shown in Fig. \ref{fig:Experimental scenarios}. Meanwhile, it
can be seen in Fig. \ref{fig:Real experimental scenarios} that the
Tx omnidirectional antenna is connected to a Rohde \& Schwarz SMBV200A
Vector Signal Generator and the Rx omnidirectional antenna is connected
to a Rohde \& Schwarz CMW500 Wideband Radio Communication Tester.
Signal carrier frequencies are set to be 0.88GHz, 2.4GHz and 2.6GHz,
respectively. Signal power decay within short distance can be reflected
using the results in Fig. \ref{fig:Experimental and fitting results}.

Fig. \ref{fig:Experimental and fitting results} shows the channel
power gain as a function of transmission distance $d$, where BPM
and unbounded pathloss model (UPM) $d^{-\alpha}$ \cite{unbounded_model_ref_3,Ref_multi_slope_1,Ref_multi_slope_2}
are applied to derive the fitting results. It can be seen that the
gaps between experimental and fitting results under UPM are large
when over small transmission distance. This is due to the singularity
of UPM at $d=0$ m and the fact that the UPM would artificially make
the received signal power greater than the transmitted signal power
when $d\in\left[0,1\right]$m\footnote{For instance, under $d^{-\alpha}$ with $\alpha=4$, if the transmission
distance $d=0.5$, Rx power would be 16 folds of the Tx power. This
is apparently inconsistent with the actual situation.}. Instead, the channel power gain is shown to be accurately modeled
using BPM even when $d$ is small. Therefore, the experimental results
are sufficient to indicate the rationality of using BPM to model pathloss
especially within short-range regions.

In addition to the above discussion, how to characterize short-range
propagation features would significantly influence the performance
evaluation and protocol design in UDN. In particular, under BPM, we
have found that network ultra-densification would eventually drain
the spatial reuse and render network capacity diminishing to be zero.
As will be discussed later, this differs from the results in \cite{Ref_multi_slope_1,Ref_multi_slope_2}
that spatial throughput scales linearly/sublinearly with BS density
under UPM.

\subsection{From Long-range Propagation to Short-range Propagation}

The above experimental results indicate that the signal power would
decay slowly with distance within short range. In constrast, it is
evident that signal power may decay rapidly with distance when long-range
transmission occur. For this reason, to model signal propagation in
UDN, it is crucial to accurately capture the characteristics of channel
gain within different propagation regions. Combining the available
results in literature\cite{Ref_multi_slope_1,BPM_original_ref,Applying_BPM_Ref2},
we propose to use multi-slope BPM\begin{small}
\begin{equation}
g_{N}^{\mathrm{B}}\left(\left\{ \alpha_{n}\right\} _{n=0}^{n=N-1};d\right)=\eta_{n}\left(1+d^{\alpha_{n}}\right)^{-1},\:R_{n}<d\leq R_{n+1},\label{eq:multi-slope pathloss model}
\end{equation}
\end{small}where $d$ denotes the distance from Tx to Rx, $R_{n}$
denotes critical distance and $\alpha_{n}$ denotes the pathloss exponent
for $R_{n}<d\leq R_{n+1}$. Meanwhile, $\eta_{0}=1$ and $\eta_{n}=\underset{i=1}{\overset{n}{\prod}}\frac{1+R_{i}^{\alpha_{i}}}{1+R_{i}^{\alpha_{i-1}}}$
are defined to maintain continuity of (\ref{eq:multi-slope pathloss model})\footnote{Other rational BPMs could be in forms like $\left(1+d\right)^{-\alpha_{n}}$
and $\min\left(1,d^{-\alpha_{n}}\right)$. The form $\left(1+d^{\alpha_{n}}\right)^{-1}$
used in this article is a typical one, which has been widely applied
in \cite{BPM_original_ref,Applying_BPM_Ref2,Applying_BPM_Ref1}.}. As signals are attenuated faster in larger distance , $\alpha_{n}<\alpha_{n+1}$
holds. Besides, the multi-slope model in (\ref{eq:multi-slope pathloss model})
can be applied into different scenarios as well.

\textit{1) Sparse outdoor scenarios:} When $N=1$, (\ref{eq:multi-slope pathloss model})
degenerates into the\textit{ single-slope BPM}, where one pathloss
exponent is used to characterize the power decay rate in the free
space. Therefore, the model is suitable for the sparse scenarios,
where Tx's are basically located apart from Rx's.

\textit{2) Dense outdoor scenarios:} When $N=2$, (\ref{eq:multi-slope pathloss model})
degenerates into the \textit{dual-slope BPM}, where two pathloss exponents
are used within and out of the critical distance. Therefore, this
model can be applied in dense outdoor scenarios such as stadium and
open gatherings, where Tx's and Rx's are located close to each other
and there are signal LoS and reflected components.

\textit{3) Dense indoor scenarios:} When $N>2$, multiple pathloss
exponents are used to capture the discrepant power decay rates within
different transmission distances. This is especially true for the
indoor scenarios, where signals from different floors and regions
may be attenuated with different rates.

\subsection{Spatial Correlation of Wireless Channels}

Besides short-range propagation of discrepant features, another dominant
property caused by network densification is the spatial correlation
of wireless channels. In general, the LOS and reflected components
between Tx and Rx contributes to the variation of wireless channels.
When Tx's (or Rx's) are in close proximity in UDN, they may have almost
identical LOS and reflected component, and consequently, the wireless
channels between them are more likely to be spatially dependent. Taking
uplink transmissions for example, assuming the users located in proximity
(several wavelength apart) are associated with different small cell
BSs, the generated desired and interfering signals over these channels
are likely to be correlated. More specifically, if the desired signals
are of great attenuation, the interfering signals from nearby users
associated with other BSs are likely to be in deep fading as well.

On the one hand, the performance evaluation of UDN in most of the
existing literatures is based on the premise that wireless channels
are uncorrelated. Therefore, the performance of UDN may be over-estimated
and is to be further investigated. On the other hand, channel independence
is the pre-assumption of most multi-antenna communications techniques,
such as coordinated multipoint (CoMP) and interference alignment (IA),
etc. For this reason, great challenges are posed towards the design
and application of these techniques, the detail of which will be discussed
later.

\begin{figure*}[t]
\begin{centering}
\includegraphics[width=7in]{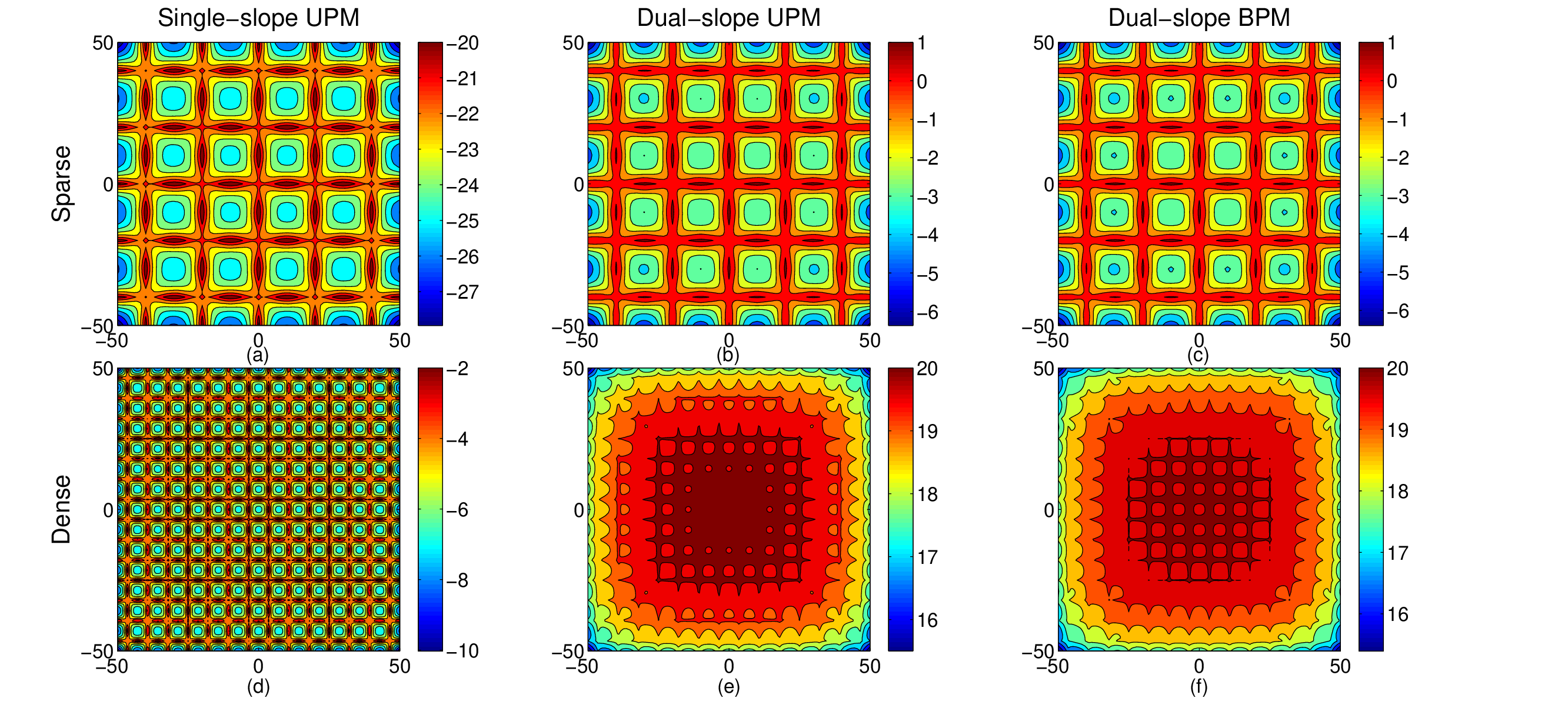}
\par\end{centering}
\caption{\label{fig:Interference distribution}. The CDF of interference under
UPM and BPM in sparse and dense deployment scenarios. The tested user
is located at the center of a square with side length 1000m. A full
spectrum reuse policy is adopted by the interferers, which are randomly
located within the scenario. Channel power gain consists of pathloss
and small scale fading. In particular, dual-slope UPM and BPM, with
$\alpha_{0}=2$, $\alpha_{1}=4$ and $R_{1}=10$m, serve as the pathloss
models. Denoting $d_{ij}$ as the distance from $\mathrm{BS}_{i}$
and $\mathrm{BS}_{j}$, $\exp\left(-d_{ij}/D_{\mathrm{C}}\right)$
is applied to quantify the correlation coefficient between $\mathrm{BS}_{i}$
and $\mathrm{BS}_{j}$, where $D_{\mathrm{C}}$ denotes the decorrelation
distance. In simulations, the transmit power of each interferer is
set to be 23dBmW and the noise power is -110 dBmW.}
\end{figure*}

\section{Interpretation of Network Densification: From the Spatial Throughput
Perspective}

Although ultra-densification is a growing trend for the future wireless
networks, there is still no consensus on how dense is ultra-dense.
To answer the question, in this part, we show some of our recent results
on the spatial throughput of wireless networks and give the interpretation
of ultra-densification from the perspective of spatial throughput
scaling law.

We define spatial throughput as follows:
\begin{align}
\mathcal{ST} & =\mu\mathbb{P}\left(\mathtt{SINR}>\text{\ensuremath{\tau}}\right)\log\left(1+\tau\right),\:\left[\mathrm{bits}/\left(\mathrm{s\cdot Hz\cdot m^{2}}\right)\right]\label{eq:spatial throughput}
\end{align}
where $\mu$ denotes the density of active links, $\tau$ denotes
the signal-to-noise-and-interference ratio (SINR) threshold and $\mathbb{P}\left(\mathtt{SINR}>\text{\ensuremath{\tau}}\right)$
denotes the success probability of data transmissions.

Intuitively, interference may degrade the SINR and the corresponding
transmission success probability especially when $\mu$ is large,
thereby serving as a limiting factor to spatial throughput. Meanwhile,
considering short-range propagation, the interference distribution
becomes complicated. Therefore, we first look into the features of
interference in wireless networks by comparing sparse and dense scenarios.

\subsection{Interference Distribution in Wireless Networks}

Fig. \ref{fig:Interference distribution} shows the cumulative distribution
function (CDF) of interference levels (in dBm) suffered by a tested
user, which is located at the center of the testing scenario. It is
shown from Fig. a that the interference CDF is almost identical under
.... when ... However, when the BS density grows, we notice that the
difference in the interference distribution is evident under different
channel models. Notably, the CDF experiences a heavy tail under UPM,
which is physically unreasonbly. This is due to the singularity of
UPM when the transmission distance approaches zero. For this reason,
it is no longer reasonable to use UPM to model pathloss under dense
deployment. Besides, to highlight the impact of channel correlation
on the interference distribution, we compare the CDFs of interference
levels with and without channel correlations under BPM in Fig. b.
Likewise, it can be seen that the influence of channel correlation
on interference distribution is negligible when interferers are sparsely
distributed. Nevertheless, the influence becomes evident when interferers
are fully densified. Specifically, we can see from Fig. b that channel
correlation would result in the degradation of interference levels.
Therefore, without considering channel correlation in UDN, the performance
of wireless networks would be greatly over-estimated, the details
of which are illustrated in the following.

\subsection{Interpret Ultra-Densification From Spatial Throughput Perspective}

\begin{figure}[t]
\begin{centering}
\includegraphics[width=3.5in]{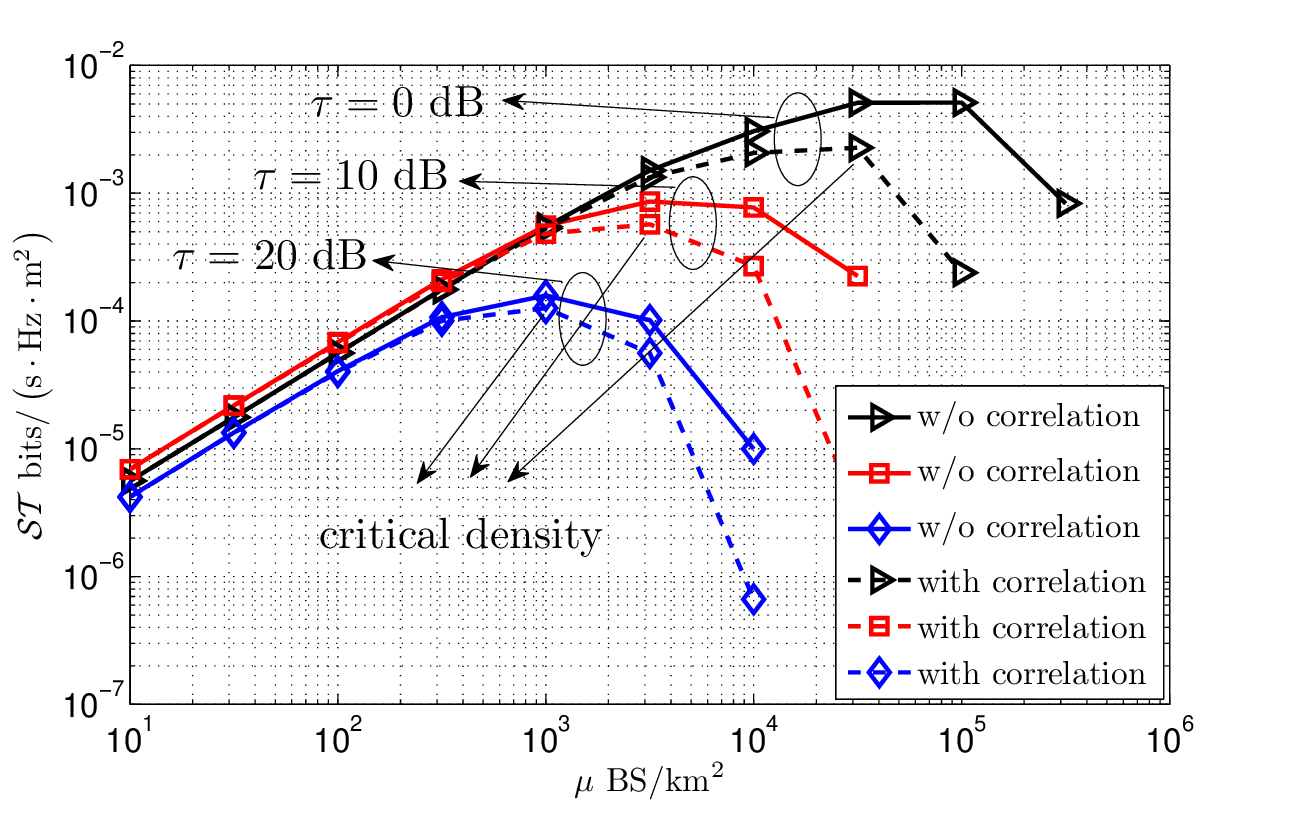}
\par\end{centering}
\caption{\label{fig:Spatial-throughput-scaling}Spatial throughput scaling
with BS density $\mu$ under different decoding threshold $\tau$
in downlink cellular networks. Single-antenna downlink users are connected
to the geometrically nearest single-antenna BSs, both of which are
assumed to be distributed according to Poisson point processes. One
user is served by one BS and no interference management techniques
are applied. Meanwhile, user density is much larger than the BS density
to guarantee that each BS is associated with at least one user. Channel
correlation coefficient is quatified by $K_{i,j}=\exp\left(-d_{i,j}/D_{\mathrm{C}}\right)$,
where $d_{i,j}$ denote the distance from $\mathrm{BS}_{i}$ to $\mathrm{BS}_{j}$
and $D_{\mathrm{C}}$ denotes the decorrelation distance. Dual-slope
BPM, i.e., $N=2$ in (\ref{eq:multi-slope pathloss model}), serves
as the pathloss model. System parameters are set as $\alpha_{0}=2$,
$\alpha_{1}=4$, $R_{\mathrm{C}}=10$m and $D_{\mathrm{C}}=10$m.}
\end{figure}

Based on (\ref{eq:spatial throughput}), we investigate the spatial
throughput of an outdoor downlink cellular network. Specifically,
we plot the spatial throughput defined in (\ref{eq:spatial throughput})
as a function of BS density under different decoding threshold $\tau$
in Fig. \ref{fig:Spatial-throughput-scaling}. To highlight the impact
of channel correlation on the spatial throughput, we compare the cases
with and without considering channel correlation. It is evident that
the gaps between the two cases are enlarged as the network is densified.
This indicates that the network performance is increasingly over-estimated
when channel correlation is not taken into account. Moreover, it is
observed in Fig. \ref{fig:Spatial-throughput-scaling} that the spatial
throughput derived under BPM scales with rate $\mu e^{-\kappa\mu}$,
i.e., first increases with $\mu$ and then decreases with $\mu$ \cite{Single_Slope_BPM_ref}.
Note that $\kappa$ is a function of system parameters. In other words,
network densification would degrade spatial throughput when BS density
is sufficiently large. For this reason, we characterize ultra-densification
in terms of the spatial throughput scaling behavior.

\textit{Ultra-densification}: Wireless networks are considered \textit{ultra-dense}
when the network density is larger than the \textit{critical density},
beyond which system spatial throughput begins to decay. The network
density has different interpretations in different network architectures.
For instance, it refers to BS density in cellular uplink/downlink
networks, while refers to density of activated Tx's in ad hoc/D2D
networks.

\begin{table}[t]
\begin{centering}
\begin{tabular}{|c|c|c|c|}
\hline
SINR Threshold & $\mu_{1}^{*}\:\left(\mathrm{BS}/\mathrm{km}^{2}\right)$ & $\mu_{2}^{*}\:\left(\mathrm{BS}/\mathrm{km}^{2}\right)$ & $\mu_{3}^{*}\:\left(\mathrm{BS}/\mathrm{km}^{2}\right)$\tabularnewline
$\tau$ $\left(\mathrm{dB}\right)$ & $\alpha_{0}=2$ & $\alpha_{0}=2$ & $\alpha_{0}=2$\tabularnewline
 & $\alpha_{1}=3$ & $\alpha_{1}=3.5$ & $\alpha_{1}=4$\tabularnewline
\hline
\hline
0 & 2.0$\times10^{5}$ & 2.51$\times10^{5}$ & 3.16$\times10^{5}$\tabularnewline
\hline
5 & 7.94$\times10^{4}$ & 1.26$\times10^{5}$ & 1.59$\times10^{5}$\tabularnewline
\hline
10 & 3.16$\times10^{4}$ & 6.31$\times10^{4}$ & 7.94$\times10^{4}$\tabularnewline
\hline
15 & 1.58$\times10^{4}$ & 3.16$\times10^{4}$ & 5.01$\times10^{4}$\tabularnewline
\hline
20 & 6.3$\times10^{3}$ & 1.58$\times10^{4}$ & 2.51$\times10^{4}$\tabularnewline
\hline
\end{tabular}
\par\end{centering}
\vspace{0.2cm}

Table II. \label{Table 1-1}Critical density $\mu^{*}$ for empirical
pathloss exponent settings \cite{Winner_II_Model_Ref}. Note that
dual-slope BPM is applied as the pathloss model, which is obtained
by setting $N=2$ in (\ref{eq:multi-slope pathloss model}).
\end{table}

Table II shows the critical densities under typical system settings,
which are derived by making an extension of the results in \cite{Single_Slope_BPM_ref}.
Under the dual-slope BPM with $\alpha_{1}=3.5$, we observe that up
to 6.31$\times10^{4}$ BSs can be deployed per square kilometer to
maximize the spatial throughput when $\tau=10\:\mathrm{dB}$. Otherwise,
if more BSs are deployed, the detriment of resulting inter-cell interference
overwhelms the benefits of spatial reuse, which degrades spatial throughput.
In this case, provided that 1 million connections/$\mathrm{km}^{2}$
are to be supported in the places such as open gathering \cite{5G_Requirement_Ref},
almost 16 connections are served by each BS. Meanwhile, it is also
observed that if users demand for higher transmission rate or equivalently
greater SINR threshold, the critical BS density is reduced. This is
because the transmission with higher rate is vulnerable and more likely
to be interrupted by inter-cell interference. In addition, Table II
indicates that larger pathloss exponents lead to larger critical densities.
In particular, the critical density increases to 2.51$\times10^{4}$
$\mathrm{BS}/\mathrm{km}^{2}$ under the dual-slope BPM with $\alpha_{1}=4$
and $\tau=20\:\mathrm{dB}$. The reason is that interference would
decay more quickly with distance under larger pathloss exponents.
As a result, its influence on spatial throughput is weakened. Note
that the empirical pathloss exponents are basically large in urban
areas, where the building blocks either form a regular Manhattan type
of grid or have more irregular locations \cite{Winner_II_Model_Ref}.

The above results also confirm the importance of applying multi-slope
BPM to model the pathloss in ultra-dense scenarios. Despite capturing
the different power decay rates over different distances, multi-slope
UPM fails to capture the power loss within short-range regions. This
makes the increase of aggregate interference power be counter-balanced
by the increase of the desired signal power. Consequently, the successful
probability in (\ref{eq:spatial throughput}) is over-estimated and
spatial throughput linearly/sublinearly increases with the BS density
$\mu$. Instead, the multi-slope BPM is capable of characterizing
moderate power decay within short-range region, which is consistent
with practice (see Fig. \ref{fig:Experimental results}). Under BPM,
the spatial throughput is shown to be eventually degraded by the over-deployment
of BSs. Therefore, multi-slope BPM could serve as a reasonable pathloss
model especially in UDN.

Aided by the interpretation of ultra-densification in Fig. \ref{fig:Spatial-throughput-scaling},
we have been fully aware of the fact that over-deployment of BSs/APs
and over-activation of devices indeed degenerate the performance of
wireless networks. Then, the following question comes up: how to breakthrough
the limitation of network densification and further enhance the capacity
of UDN? The detail will be discussed in the following section.

\section{Possible Approaches and Challenges to Improve Capacity in UDN}

For capacity enhancement, a number of state-of-the-art techniques
can be applied. However, challenges exist as well in practical implementations
due to the channel and interference characteristics in UDN.

\subsection{Interference Management}

Since severe and complicated interference is the key factor to bottleneck
the capacity in UDN, interference management techniques are of the
greatest potential to combat interference and improve network capacity.
Interference cancellation and interference coordination are two prevalent
interference management techniques. Applying interference cancellation,
interference signals are rebuilt, decoded and finally successively
or parallelly removed from the aliasing signals until the desired
signal is retrieved. Decoding interference signals requires that interference
signals are of great disparity. For interference coordination, with
the aid of multi-antenna technologies, desired signal and interference
signals are forced to be spatially orthogonal at Rx's via jointly
designing precoders by multiple Tx's (or both Tx's and Rx's). Channel
independence is the premise for joint precoder design. However, the
interference and channel features in UDN greatly degrade the performance
of interference cancellation and interference coordination. On the
one hand, interference signals basically stem from the interferers,
which are geometrically close to the intended Rx in dense scenarios.
Accordingly, the interference levels become less divergent. On the
other hand, as discussed earlier, network densification makes the
channels of transmission pairs in close proximity become spatially
correlated as well. This may directly ruin the merits brought by multi-antenna
based interference management techniques.

\subsection{Non-Orthogonal Multiple Access}

NOMA also serves as a promising method to improve user connectivity
and network capacity by fully multiplexing available spectrum resources
via non-orthogonal spectrum sharing \cite{Ref_NOMA}. The key to NOMA
is to cancel the proactively introduced spectrum-domain interference
in other domains. For instance, the power domain based NOMA (PD-NOMA)
and sparse code multiple access (SCMA) exploit the degree of freedom
in power domain and code domain, respectively. However, as discussed
earlier, as interference cancellation would lose the merit due to
the narrowed interference levels, the performance of NOMA is degrades
in UDN. Worsestill, NOMA is basically co-designed with resource allocation.
To enable effcient resource allocation, accurate overhead for channel
estimation and information exchange is required. Yet, the overhead
caused by network over-deployment would be massive and further increase
with the user density, which in turn ruins the potential benefits
of NOMA. Hence, how to design scalable NOMA remains to be an open
issue in UDN.

\subsection{Millimeter-Wave Communications}

The reduced transmission distance makes it possible to apply millimeter-wave
(mm-Wave) communications in UDN \cite{Ref_mm_Wave}. Under mm-Wave
bands over 30-100GHz, higher data rates and larger network capacity
can be readily guaranteed. Moreover, interference signals are more
rapidly decays with transmission distance over higher frequency bands,
which ensures lower interference levels in mm-Wave communications.
Even when interferers are in close proximity, interference could be
spatially avoided with the aid of directional antennas. Despite the
benefits, antenna directivity feature would result in serious problems
as well. Especially, the antenna directions, which depend on the relative
locations of Tx's and the intended Rx's, have to adjust dynamically
when Tx's or Rx's are on the move. On the one hand, acquiring instantaneous
location information in UDN is a waste of overhead, which limits the
implementation of mm-Wave communications. On the other hand, the adjustment
of antenna directions among adjecent transmission pairs would make
the transmission pairs, which are interference-free, potentially interfere
with each other. Consequently, how to design mm-Wave techniques under
the above considerations is challenging.

\section{Conclusion Remarks}

As an inevitable tendency in future wireless networks, network densification
significantly reduces transmission distance and make signal propagation
transit from long- to short-range propagation. In this article, we
discuss the impact of short-range propagation on the wireless network
performance, especially considering densely deployed scenarios. Remarkably,
we found that network densification would eventually drain the spatial
resources and degrade network performance when network density exceeds
a critical density. More importantly, aided by the critical density,
we interpret network ultra-densification from the perspective of spatial
throughput scaling law. The result serves as a guidance for network
deployment, as it indicates under what circumstances deploying more
BSs/APs is beneficial to enhancing network capacity. In summary, this
article has merely shed light on a drop in the bucket of UDN. It is
imperative to fully understand and exploit the inherent features of
UDN, thereby achieving the aggressive goals of future wireless networks.

\bibliographystyle{IEEEtran}
\bibliography{ref_magazine}

\begin{thebibliography}{10}
\providecommand{\url}[1]{#1}
\csname url@samestyle\endcsname
\providecommand{\newblock}{\relax}
\providecommand{\bibinfo}[2]{#2}
\providecommand{\BIBentrySTDinterwordspacing}{\spaceskip=0pt\relax}
\providecommand{\BIBentryALTinterwordstretchfactor}{4}
\providecommand{\BIBentryALTinterwordspacing}{\spaceskip=\fontdimen2\font plus
\BIBentryALTinterwordstretchfactor\fontdimen3\font minus
  \fontdimen4\font\relax}
\providecommand{\BIBforeignlanguage}[2]{{%
\expandafter\ifx\csname l@#1\endcsname\relax
\typeout{** WARNING: IEEEtran.bst: No hyphenation pattern has been}%
\typeout{** loaded for the language `#1'. Using the pattern for}%
\typeout{** the default language instead.}%
\else
\language=\csname l@#1\endcsname
\fi
#2}}
\providecommand{\BIBdecl}{\relax}
\BIBdecl

\bibitem{5G_Requirement_Ref}
\BIBentryALTinterwordspacing
{IMT-2020 (5G) Promotion Group}, ``{5G} vision and requirements,'' Tech. Rep.,
  Dec. 2015. [Online]. Available:
  \url{http://www.imt-2020.org.cn/en/documents/download/3}
\BIBentrySTDinterwordspacing

\bibitem{UDN_benefit_ref}
W.~Webb, \emph{ArrayComm}.\hskip 1em plus 0.5em minus 0.4em\relax London, U.K.:
  Ofcom, 2007.

\bibitem{Architecture_Ref}
P.~K. Agyapong, M.~Iwamura, D.~Staehle, W.~Kiess, and A.~Benjebbour, ``Design
  considerations for a {5G} network architecture,'' \emph{IEEE Commun. Mag.},
  vol.~52, no.~11, pp. 65--75, Nov. 2014.

\bibitem{User_Centric_UDN_Ref}
S.~Chen, F.~Qin, B.~Hu, X.~Li, and Z.~Chen, ``User-centric ultra-dense networks
  for {5G}: challenges, methodologies, and directions,'' \emph{IEEE Wireless
  Commun.}, vol.~23, no.~2, pp. 78--85, April 2016.

\bibitem{Ref_multi_slope_1}
X.~Zhang and J.~G. Andrews, ``Downlink cellular network analysis with
  multi-slope path loss models,'' \emph{IEEE Trans. Commun.}, vol.~63, no.~5,
  pp. 1881--1894, May 2015.

\bibitem{Ref_multi_slope_2}
M.~Ding, P.~Wang, D.~Lopez-Perez, G.~Mao, and Z.~Lin, ``Performance impact of
  {LoS} and {NLoS} transmissions in dense cellular networks,'' \emph{IEEE
  Trans. Wireless Commun.}, vol.~15, no.~3, pp. 2365--2380, Mar. 2016.

\bibitem{Resource_allocation_Ref}
Z.~Zhou, M.~Dong, K.~Ota, and Z.~Chang, ``Energy-efficient context-aware
  matching for resource allocation in ultra-dense small cells,'' \emph{IEEE
  Access}, vol.~3, pp. 1849--1860, Sep. 2015.

\bibitem{unbounded_model_ref_3}
H.~S. Dhillon, R.~K. Ganti, F.~Baccelli, and J.~G. Andrews, ``Modeling and
  analysis of {K-Tier} downlink heterogeneous cellular networks,'' \emph{IEEE
  J. Sel. Areas Commun.}, vol.~30, no.~3, pp. 550--560, Apr. 2012.

\bibitem{BPM_original_ref}
H.~Inaltekin, M.~Chiang, H.~V. Poor, and S.~B. Wicker, ``On unbounded path-loss
  models: effects of singularity on wireless network performance,'' \emph{IEEE
  J. Sel. Areas Commun.}, vol.~27, no.~7, pp. 1078--1092, Sep. 2009.

\bibitem{Applying_BPM_Ref2}
R.~K. Ganti and M.~Haenggi, ``Interference and outage in clustered wireless ad
  hoc networks,'' \emph{IEEE Trans. Inf. Theory}, vol.~55, no.~9, pp.
  4067--4086, Sep. 2009.

\bibitem{Applying_BPM_Ref1}
H.~Inaltekin, ``Gaussian approximation for the wireless multi-access
  interference distribution,'' \emph{IEEE Trans. Signal Process.}, vol.~60,
  no.~11, pp. 6114--6120, Nov. 2012.

\bibitem{Single_Slope_BPM_ref}
J.~Liu, M.~Sheng, L.~Liu, and J.~Li, ``Effect of densification on cellular
  network performance with bounded pathloss model,'' \emph{IEEE Commun. Lett.},
  vol.~21, no.~2, pp. 346--349, 2017.

\bibitem{Winner_II_Model_Ref}
P.~Ky{\"o}sti, J.~Meinil{\"a}, L.~Hentil{\"a}, X.~Zhao, T.~J{\"a}ms{\"a},
  C.~Schneider, M.~Narandzic, M.~Milojevic, A.~Hong, J.~Ylitalo \emph{et~al.},
  ``Winner {II} channel models,'' \emph{WINER II Public Deliverable}, pp.
  42--44, 2007.

\bibitem{Ref_NOMA}
L.~Dai, B.~Wang, Y.~Yuan, S.~Han, C.~l.~I, and Z.~Wang, ``Non-orthogonal
  multiple access for {5G}: solutions, challenges, opportunities, and future
  research trends,'' \emph{IEEE Commun. Mag.}, vol.~53, no.~9, pp. 74--81,
  Sept. 2015.

\bibitem{Ref_mm_Wave}
R.~Baldemair, T.~Irnich, K.~Balachandran, E.~Dahlman, G.~Mildh, Y.~Selén,
  S.~Parkvall, M.~Meyer, and A.~Osseiran, ``Ultra-dense networks in
  millimeter-wave frequencies,'' \emph{IEEE Commun. Mag.}, vol.~53, no.~1, pp.
  202--208, Jan. 2015.

\end{thebibliography}

\end{document}